\documentclass[conference, a4paper]{IEEEtran}
\usepackage{times}

\usepackage{graphicx}
\usepackage{cite}
\usepackage{amsmath}
\begin{document}

\title{Random Matrix Theory of Resonances: an Overview}

\author{\IEEEauthorblockN{%
    Yan V. Fyodorov\IEEEauthorrefmark{1}}
\medskip
  \IEEEauthorblockA{\IEEEauthorrefmark{1}
    King's College London, Department of Mathematics, London  WC2R 2LS, United Kingdom\\
    e-mail: yan.fyodorov@kcl.ac.uk}}


\maketitle

\begin{abstract}
Scattering of electromagnetic waves in billiard-like systems has become a standard experimental tool of studying properties associated with Quantum Chaos.  Random Matrix Theory (RMT) describing statistics of eigenfrequencies and associated eigenfunctions remains one of the pillars of theoretical understanding of quantum chaotic systems.  In a scattering system coupling to continuum via antennae converts real eigenfrequencies into  poles of the scattering matrix in the complex frequency plane and the associated eigenfunctions into decaying resonance states. Understanding statistics of these poles, as well as associated non-orthogonal resonance eigenfunctions within RMT approach is still possible, though much more challenging task.

\end{abstract}

\section{Introduction and formalism}
Scattering of electromagnetic waves in billiard-like microwave resonators or, in optical range, dielectric microcavities, has become a standard experimental tool of studying properties associated with Quantum Chaos, see e.g. reviews \cite{kuhl13,grad14,diet15,hcao15}. Random Matrix Theory (RMT) describing statistics of eigenfrequencies (referred to as "energy levels" in the quantum mechanical context, the terminology to be used in the rest of the review) and associated eigenfunctions is one of the pillars of theoretical understanding of quantum chaotic systems. Scattering set up drastically changes the system properties by coupling to the continuum via antennae, thereby converting discrete energy levels into decaying resonance states. Such states can be associated with poles of the scattering matrix in the complex energy plane. Understanding statistics of such poles, as well as associated residues related to non-orthogonal eigenfunctions within RMT approach is still possible, but in full generality remains a challenging task for theoreticians.
This presentation aims at giving a short overview of present knowledge in this field.

The starting point is an RMT model introduced in \cite{verb85} and described in detail in \cite{fyod97}, and more recently in \cite{fyod05,fyod11}. The model deals with the unitary $M\times M$ energy-dependent scattering matrix $S(E)$ for quantum-chaotic system with $M$ open channels
\begin{equation}\label{1}
S(E)=\frac{{\bf 1}-iK}{{\bf 1}+iK}, \quad\quad\mbox{ where } \quad \quad K=W^{\dagger}\frac{1}{E-H}W
\end{equation}
with $N\times N$ , $N\gg 1$  random Gaussian (real symmetric GOE, $\beta=1$, Hermitian GUE, $\beta=2$,  or real quaternion GSE, $\beta=4$) matrix  $H$  used to model spectral properties of the Hamiltonian of closed system of quantum-chaotic nature.  The columns of $N\times M$ matrix $W$ of  {\it coupling amplitudes} to $M$ open scattering channels can be taken either as  fixed orthogonal vectors satisfying $\sum_{i=1}^N \overline{W}_{ai}W_{bi}=\gamma_a\delta_{ab}, \, \gamma_a>0 \,\, \forall a=1,\ldots, M $
or alternatively the entries are chosen to be independent random Gaussian \cite{soko89} $: \quad\left\langle \overline{W}_{ai}W_{bj}\right\rangle=N^{-1}\gamma_a\delta_{ab}\delta_{ij}, \quad i, j=1,\ldots, N $, with angular brackets standing for the ensemble averaging. Equivalently, entries $S_{ab}(E)$ of the scattering matrix can be rewritten as \cite{verb85}
\begin{equation}\label{2}
S_{ab}(E)=\delta_{ab}-2i\sum_{ij} W^*_{ai}\left[\frac{1}{E-{\cal H}_{eff}}\right]_{ij}W_{jb},
\end{equation}
with an effective non-Hermitian Hamiltonian
\begin{equation}\label{3}
 {\cal H}_{eff}=H-i\Gamma, \, \quad \, \Gamma=WW^{\dagger}\ge 0
\end{equation}
whose $N$ complex eigenvalues $z_n=E_n-i\frac{\Gamma_n}{2}$ provide poles of the scattering matrix in the complex energy plane, commonly referred to as the {\it resonances}. The goal is to describe the statistics of positions of these poles in the complex plane, as well as statistics of the associated residues
related to non-orthogonal eigenvectors of non-Hermitian Hamiltonian ${\cal H}_{eff}$, using the random matrix statistics for $H$ as an input.

\section{Statistics of S-matrix poles in some limiting cases.}
\subsection{Weak-coupling regime and Porter-Thomas distribution}
When $W\to 0$ the anti-Hermitian part  $-iWW^{\dagger}$ can be treated as a perturbation of the Hermitian part $H$. The latter matrix is characterized by real eigenvalues $E_n$ and orthonormal eigenvectors $\left.|n\right\rangle$ which are random vectors uniformly spanning the $N-$dimensional unit sphere.
 In this regime one expects the resonances to be largely {\it non-overlapping}, that is $\Gamma_n \ll \Delta$, where $\Delta$ is the mean spacing between neighbouring real eigenvalues for  $H$. To the first order
\[E_n\to z_n=E_n-i\frac{\Gamma_n}{2}, \quad \quad \mbox{with}\quad \quad \Gamma_n=2\left\langle n|WW^{\dagger}|n\right\rangle \]
implying for $M$ equivalent channels with coupling constants $\gamma_a=\gamma\ll 1$ the $\chi^2$ distribution of scaled resonance widths $y_n=\pi\Gamma_n/\Delta\ll 1$ \cite{port56}:
\begin{equation} \label{PT} {\cal P}_M^{(\beta)}(y)=\frac{(\beta/2)^{M\beta/2}}{\gamma\Gamma(M\beta/2)}
\left(\frac{y}{\gamma}\right)^{\frac{M\beta}{2}-1} e^{-\frac{\beta y}{4\gamma}}
\end{equation}
This expression known as the {\it Porter-Thomas} distribution favourably agrees with a lot of experimental data in billiards, from \cite{alt95} to  recent experiments \cite{difa12} in a stadium
billiard embedded into a two dimensional photonic crystal realized on a silicon-on-insulator substrate. Note however recently reported
deviations in high-precision neutron scattering \cite{koeh10}, see also discussion below and references in \cite{fyod15}.

\subsection{Limiting case of very many open channels}
On the other hand when couplings $\gamma>0$ are fixed and the number $M$ of open channels is very large and comparable with the number $N\gg 1$ of internal states we expect that typically resonances {\it  overlap strongly}:  $\Gamma_n\gg \Delta$. The mean density of complex eigenvalues of  ${\cal H}_{eff}=H-i\Gamma, \, \, \Gamma=WW^{\dagger}$ for many channels  $0<m=M/N<1$
was found analytically in \cite{haak92,lehm95a}. Generically, the density of resonances has the form of a cloud separated by
a  gap from the real axis. For larger couplings $\gamma$ second cloud emerges, as depicted in Fig.1 taken from \cite{lehm95a}.
\begin{figure}[h!]
    \begin{center}
   \includegraphics[scale=0.5]{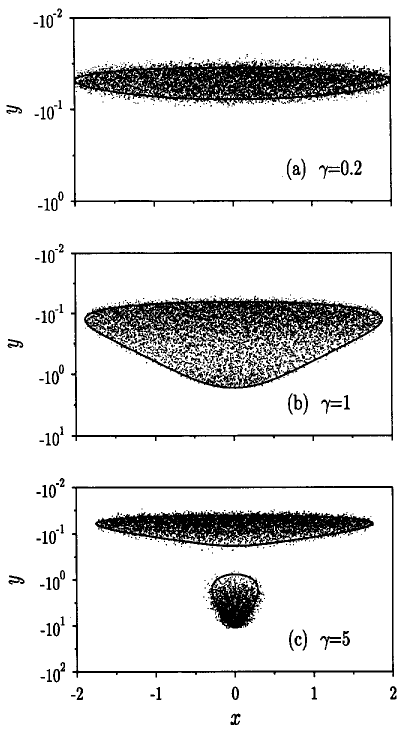}
    \end{center}
\vspace{-5pt}
    \caption
{\small The density of resonances in RMT model with many open channels, from \cite{lehm95a}. }
\end{figure}
The gap has the physical meaning of a correlation length in energy-dependent scattering observables, see \cite{lehm95a} and also matches semiclassical considerations \cite{gasp93} and is observed experimentally in microwave billiards \cite{bark13}.

\section{Non-perturbative results for statistics of S-matrix poles}

The first systematic non-perturbative investigation of resonances within RMT was undertaken by Sokolov $\&$ Zelevinsky \cite{soko89} for the special case $M=1$. They provided an explicit expression for the joint probability density of positions $z_1,\ldots , z_N$ of all $N$ eigenvalues of ${\cal H}_{eff}=H-iWW^{\dagger}$ in the complex energy plane assuming random Gaussian coupling amplitudes $\left\langle W_iW_j\right\rangle=\frac{\gamma}{N}$ with fixed $\gamma>0$. For $\beta=1$  they found

\vspace{-0.3cm}

\[
{\cal P}_{M=1}^{(\beta=1)}(z_1,\ldots , z_N)\propto \prod_{k=1}
\frac{e^{-\frac{N}{\gamma}\mbox{\small Im}z_k}}{\sqrt{\gamma \mbox{\small Im}z_k}} \prod_{i<j}\frac{|z_i-z_j|^2}{|z_i-\overline{z}_j|}
\]


\begin{equation} \label{3}
\times e^{-\frac{N}{4}\sum_{k=1}^N(\mbox{\small Re}z_k )^2-\frac{N}{2}\sum_{k<l}^N\mbox{\small Im}z_k \mbox{\small Im}z_l}
\end{equation}
This density was then used to predict that for $\gamma$ large enough the resonances "reorganize" into one short-lived resonances and a cloud of $N-1$ long lived ones (analogue of {\it Dicke superradiance}), see recent optical experiments in \cite{liu14}.

Similar, but somewhat simpler expression can be derived for $\beta=2$. In that case the limiting distribution of the resonance widths $\Gamma_n$ can be
derived non-perturbatively for any fixed number of channels $M\ll N\to \infty $ \cite{fyod96}. To that end
 one defines  "renormalized coupling strengths"  $g_a=\frac{1}{2}\left(\gamma_c+\frac{1}{\gamma_c}\right)$ for all channels $a=1,\ldots,M$.
 Then
the probability density of the scaled resonance widths $y_n=\pi\Gamma_n/\Delta$ is given for $M$ equivalent channels with $g_1=\ldots =g_M\equiv g$ by \cite{fyod96}
\begin{equation}\label{denbeta2}
{\cal P}_M^{(\beta)}(y)=\frac{(-1)^M}{(M-1)!}y^{M-1}\frac{d^M}{dy^M}
\left\{e^{-y g}\left(\frac{\sinh{y}}{y}\right)\right\}
\end{equation}
 For weak coupling $\gamma\ll 1$ we have $g \sim \gamma^{-1}\gg1 $ hence typically $y\sim g^{-1}\ll 1$
 and we are back to the Porter-Thomas distribution (\ref{PT}).
 In contrast, for the  {\it perfect coupling} case $g=1$  the power-law tail emerges ${\cal P}_M^{(\beta)}(y)\propto 1/y^2$ so that some resonances may  overlap strongly. This favourably agrees with the numerics for quantum chaotic graphs \cite{kott00}, see Fig.2.
\begin{figure}[h!]
    \begin{center}
   \includegraphics[scale=0.35]{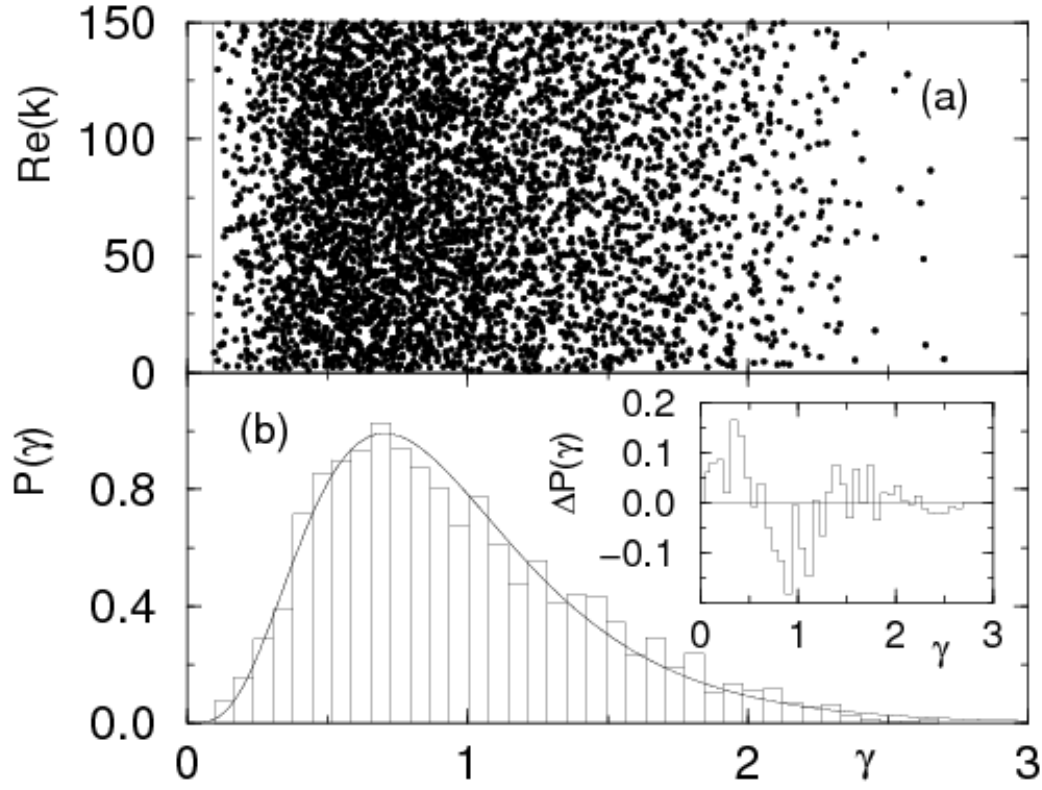}
    \end{center}
\vspace{-5pt}
    \caption
{\small Top: 5000 resonances for a single realization of $\beta=2$ chaotic scattering in quantum graphs ( taken from T. Kottos and U. Smilansky \cite{kott00}). Bottom: Resonance widths distribution as compared with the RMT analytical prediction (\ref{denbeta2}).}
\end{figure}
In fact not only the mean density of resonances, but also all higher correlation functions can be found explicitly \cite{fyod99}
as resonances form for $\beta=2$ a {\it determinantal  process} in the complex plane.
For $\beta=1$ only the mean density of resonances in the complex plane is known explicitly for $1\le M<\infty$  \cite{somm99} and is given by a rather complicated expression. E.g. in the simplest case $M=1$ the probability density of the scaled resonance widths $y_n$ is given by
\begin{equation}
{\cal P}_{M=1}^{(\beta=1)}(y)= \frac{1}{4\pi}\frac{d^2}{dy^2}
\int_{-1}^{1}(1-\lambda^2)e^{2\lambda y}(g-\lambda){\cal F}(\lambda,y)\,d\lambda
\end{equation}
where
\begin{equation}
{\cal F}(\lambda,y)=\int_g^{\infty}dp_1\frac{e^{-y p_1}}{(\lambda-p_1)^2\sqrt{(p_1^2-1)(p_1-g)}}
\end{equation}
\[
\times \int_1^{g}dp_2(p_1-p_2)\frac{e^{-y p_2}}{(\lambda-p_2)^2\sqrt{(p_2^2-1)(g-p_2})}
\]
That formula was confirmed in microwave experiments by Kuhl et al. \cite{kuhl08}, see Fig.3.
\begin{figure}[h!]
    \begin{center}
   \includegraphics[scale=0.6]{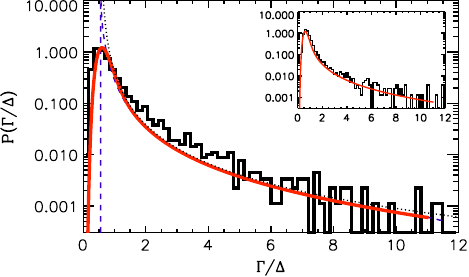}
    \end{center}
\vspace{-5pt}
    \caption
{\small Resonance widths distribution for $\beta=1, M=1$  scattering in a microwave resonator as compared with the RMT analytical prediction for perfect coupling, taken from \cite{kuhl08}.  The inset shows direct RMT simulations. The characteristic tail $1/\Gamma^2$ is well-developed.}
\vspace{-5pt}
\end{figure}

Recently reported deviations from Porter-Thomas distribution in neutron scattering by Koehler et al. \cite{koeh10}  stimulated many heuristic proposals to modify Porter-Thomas distribution beyond the small widths region.  Motivated by that we analyzed the exact formulae for $\beta=1$ in order to extract a  non-perturbative asymptotics of ${\cal P}(y)$ valid for $g \gg 1$ and any scaled widths $y$. For $M=1$ and $g\sim (2\gamma)^{-1}\gg 1$ we have found \cite{fyod15}
\begin{equation}
{\cal P}_{M=1}^{(\beta=1)}(y)= - \frac{d}{dy}\left( \Phi(y) \mathrm{erfc}\sqrt{g y/2}\right)
\end{equation}
where $\mathrm{erfc}(z)=\frac{2}{\sqrt{\pi}}\int_z^{\infty}e^{-t^2}dt$ and the correction factor
\[
\Phi(y)=\frac{1}{2}\left\{K_0\left(\frac{y}{2}\right)\left[\cosh{y}-\frac{\sinh{y}}{y}\right]
+K_1\left(\frac{y}{2}\right)\sinh{y}\right\}
\]
 For small widths  $\Phi(y\ll 1)\approx 1 $ and we are back to the Porter-Thomas distribution (\ref{PT}).


Finally, due coupling to continuum statistics of the real parts ${\mbox Re}\, z_n$ is significantly modified
in comparison with eigenfrequencies $E_n$ of the closed systems. Although no results derived from the first principles are yet available, an interesting generalization of the {\it Wigner surmise} to open system was suggested in \cite{poli12}, and compared with experimental data.

\section{Resonance Eigenfunction non-Orthogonality}
The effective non-Hermitian Hamiltonian ${\cal H}_{eff}=H-iWW^{\dagger}$
has a set of "right" $\left.|R_n\right\rangle$ and "left" $\left\langle L_n\right.|$ eigenvectors:
\begin{equation}{\cal H}_{eff}\left.|R_n\right\rangle=z_n\left.|R_n\right\rangle, \quad
\left\langle L_n\right.|{\cal H}_{eff}=z_n\left\langle \left.L_n\right|\right.\end{equation}
with  bi-orthogonality properties:
\[\left.\left\langle L_n|R_n\right\rangle\right.=\delta_{mn}, \quad   \sum_{n=1}^N \left.|R_n\right\rangle\left\langle L_n|\right.={\bf 1}\]
Note however that  $\left\langle L_n|L_m\right\rangle\ne\delta_{nm}\ne \left\langle R_n|R_m\right\rangle$ showing that the eigenmodes of open systems are
no longer orthogonal. The corresponding non-orthogonality overlap matrix
\[{\cal O}_{mn} = \left\langle L_m|L_n\right\rangle \left\langle R_m|R_n\right\rangle\]
 shows up in various physical observables of quantum chaotic systems, such as e.g. decay laws \cite{savi97}, "Petterman factors"  describing excess noise in open laser resonators \cite{scho00}, as well as in sensitivity of the resonance widths to small perturbations \cite{fyod12}. RMT can be most efficiently applied to statistics of ${\cal O}_{mn}$ in the weak coupling regime $\gamma\ll 1$ of
isolated resonances  \cite{scho00,fyod12,poli09}, or in the opposite case of strongly overlapping resonances in extremely open systems with number of open channels $M\sim N$ \cite{jani99,sant01}. Non-perturbative results for finite $M$ are scarce and available only for $\beta=2$  \cite{scho00,fyod02}.

\subsection{Resonance widths "shifts" as a signature of non-orthogonality}
Suppose we slightly perturb the scattering system: ${\cal H}_{eff}\to {\cal H}_{eff} +\alpha V$ (e.g. by moving scatterers, or billiard walls)  with
 a Hermitian $V=V^{\dagger}$ and  $\alpha\to 0$. The  width of each resonance $\Gamma_n$ changes typically linearly in the perturbation strength: $\delta\Gamma_n\propto \alpha$, and one can show that such "shifts" are non-vanishing only due to non-orthogonality of the resonance eigenffunctions \cite{fyod12}.
 In the regime of non-overlapping resonances one can further show that
\begin{equation}
\delta\Gamma_n = i\alpha \sum_{m\ne n} \frac{\left\langle m|G|m\right\rangle}{E_n-E_m},
\end{equation}
where real eigenvalues $E_n$ and orthonormal eigenvectors $\left.|n\right\rangle$ characterize the "closed" chaotic system and
\[\left.G=WW^{\dagger}|n\right\rangle\left\langle n|V+V|n\right\rangle\left\langle n|WW^{\dagger}\right.\]

One can use RMT results for $E_n$ and $\left.|n\right\rangle$  to derive the distribution of "shifts" $\delta\Gamma_n$ which shows a power-law tail \cite{fyod12} $\sim(\delta\Gamma_n)^{-(\beta+2)}$. For experimental verification see \cite{gros14}.
\begin{figure}[h]
    \begin{center}
   \includegraphics[scale=0.3]{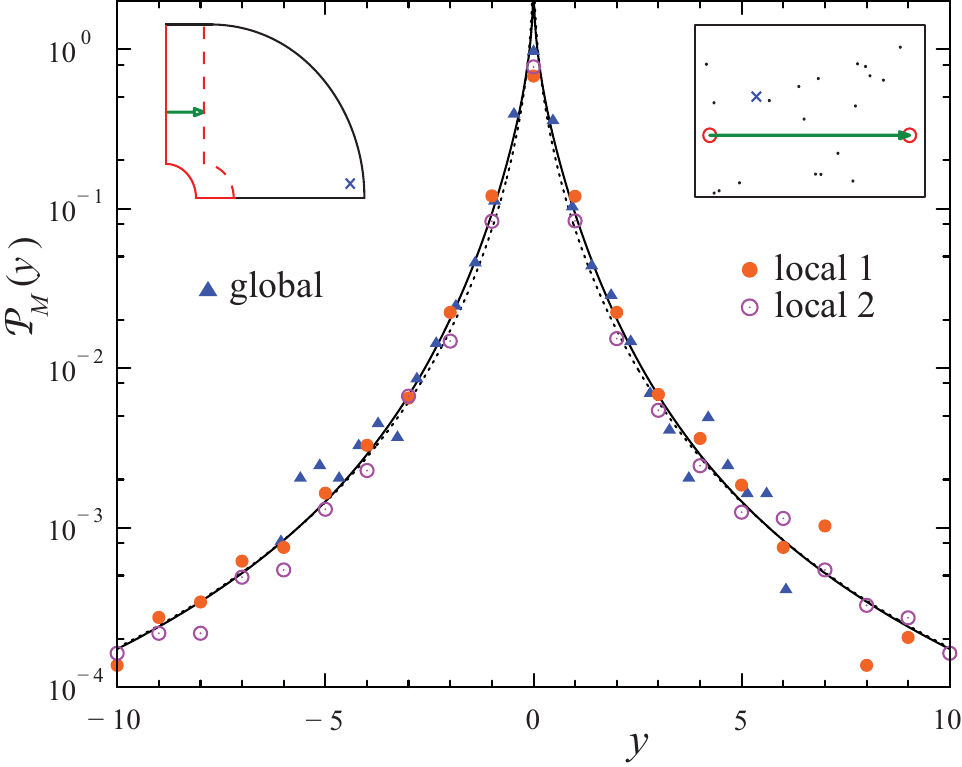}
    \end{center}
\vspace{-5pt}
    \caption
{\small Resonance widths "shifts" distribution for $\beta=1, M=1$  scattering in a microwave billiard resonator as compared with the RMT analytical prediction, for various types of perturbations (global vs. local). From  Gros et al.\cite{gros14}}
\vspace{-10pt}
\end{figure}

\section{Closely related topics.}
There are quite a few works on resonances in wave chaotic scattering related to the topic of this review but going beyond its direct remit.
Further information can be found in the references below.
\begin{itemize}
\item Resonance theory for time-periodic (Floquet) or stroboscopic systems, relations to truncations of unitary random matrices: \cite{zycz00,fyod00,gluck02,keat08,nova13}.
\item Semiclassical structure of resonances in chaotic systems and Fractal Weyl law, see \cite{luetal03} and further references in the review \cite{nova13}. For a recent experiment see \cite{Lippolis}.
\item Resonance properties in systems with diffusion of waves \cite{borg91,skip06}, with wave scattering generated by point-like active scatterers \cite{goet11}, and in interacting many-body chaotic systems \cite{cela11}.
\item Resonances in systems with Anderson localization \cite{tito00,weis06}, see also references in \cite{gure12}.
\item Finally we mention growing interest in these questions in pure mathematics community, see e.g. \cite{kozh15}.
\end{itemize}


\end{document}